\documentclass[aps,pra,twocolumn,twoside,showpacs,floatfix]{revtex4}
\usepackage{graphicx,amsmath,amssymb}
\def\PRA{{Phys. Rev.} A }
\newcommand{\traket}[1]{\mbox{\ensuremath{|#1\rangle}}}
\newcommand{\trabra}[1]{\mbox{\ensuremath{\langle#1|}}}

\begin{document}
%%%%%%%
\title{Quantum state reconstruction using binary data from on/off photodetection}
\author{Giorgio Brida, Marco Genovese, Marco Gramegna, Alice Meda, Fabrizio
Piacentini, Paolo Traina}
\affiliation{I.N.RI.M., Strada delle Cacce 91, Torino, Italia}
\author{Enrico Predazzi}
\affiliation{INFN and Dipartimento di Fisica Teorica, Universit`a di Torino, 
I-10125 Torino, Italia}
\author{Stefano Olivares, Matteo G.~A.~Paris}
\affiliation{CNISM UdR Milano Universit\`a, I-20133 Milano, Italia}
\affiliation{Dipartimento di Fisica dell'Universit\`a degli Studi di Milano, I-20133
Milano, Italia}
%%%%%%%
\begin{abstract}
The knowledge of the density matrix of a quantum state plays a fundamental
role in several fields ranging from quantum information processing to experiments
on foundations of quantum mechanics and quantum optics. Recently, a method has been
suggested and implemented in order to obtain the reconstruction of the diagonal
elements of the density matrix exploiting the information achievable with realistic
on/off detectors, e.g. silicon avalanche photo-diodes, only able to discriminate
the presence or the absence of light. The purpose of this paper is to provide
an overview of the theoretical and experimental developments of the on/off method,
including its extension to the reconstruction of the whole density matrix.
\end{abstract}
\date{\today}
%%%%%%%
\maketitle
%%%%%%%
\section{Introduction}
The knowledge of the density matrix of a quantum state is fundamental for
several applications, ranging from quantum information
\cite{Bouwmeester2000} to the foundations of quantum mechanics
\cite{MG} and quantum optics
\cite{man1,man2,man3,Mandel,smithey,dariano,raymerLNP,dariano2}.
In turn, many efforts have been devoted to find reliable methods to
fully or partially reconstruct the density matrix especially. This is especially
true for the density matrix in the photon number basis, and in this case
the reconstruction of the diagonal elements, i.e. the photon statistics,
is of great relevance for the characterization and use of the state in
quantum communication and information processing.
In fact, the field of photodetection has received much attention in the
last decades, however, the choice of a detector with internal gain suitable
for the measurement is still a non trivial task when the flux of the photons to
be counted is such that more than one photon is detected in the
time-window of the measurement, which in turn is set by the detector
pulse-response, or by an electronic gate on the detector output, or
by the duration of the light pulse. In this case, we need a
congruous linearity in the internal current amplification process:
each of the single electrons produced by the different photons in
the primary step of the detection process (either ionization or
promotion to a conduction band) must experience the same average
gain and this gain must have sufficiently low spread. The
fulfillment of both requisites is necessary for the charge integral
of the output current pulse be proportional to the number of
detected photons. Photon detectors that can operate as photon
counters are rather rare \cite{d,e}. Among these, Photo-Multiplier Tubes (PMT's)
\cite{burle,h} and hybrid photo-detectors \cite{NIST,m} have the drawback
of a low quantum efficiency, since the detection starts with the
emission of an electron from the photo-cathode. Solid state detectors
with internal gain, in which the nature of the primary detection
process ensures higher efficiency, are still under development.
Highly efficient thermal photon counters have also been used, though
their operating conditions are still extreme (cryogenic conditions)
to allow common use \cite{xxx, serg, f}. Better results can be in principle
be obtained using photon chopping in conjuction with single-photon detectors,
whereas the experimental implementation of loop-detectors has shown interesting
performances.
The advent of quantum
tomography provided an alternative method to measure photon number
distributions \cite{mun,c}. However, the tomography of a state, which
has been applied to several quantum states \cite{raymerLNP,q,r,s,t}, needs
the implementation of homodyne detection, which in turn requires the
appropriate mode matching of the signal with a suitable local
oscillator at a beam splitter. Such mode matching is a particularly
challenging task in the case of pulsed optical fields.
\par
Photodetectors that are usually employed in quantum optics such as
Avalanche Photo-Diodes (APD's) operating in the Geiger mode
\cite{rev, serg,i,l} appears, at a fist sight, to be definitely
useless as photon counters. They are the solid state photo-detectors
with the highest quantum efficiency and the greatest stability of
the internal gain. However, they have the obvious drawback that the
breakdown current is independent of the number of detected photons,
which in turn cannot be determined. The outcome of these APD's is
either "off" (no photons detected) or "on" {\em i.e.} a click
indicating the detection of one or more photons. Actually, such an
outcome can be provided by any photodetector (PMT, hybrid
photodetector, cryogenic thermal detector) for which the charge
contained in dark pulses is below that of the output current pulses
corresponding to the detection of at least one photon. Notice that
for most high-gain PMT's the anodic pulses corresponding to no
photons detected can be easily discriminated by a threshold from
those corresponding to the detection of one or more photons. Despite
the above considerations, a certain effort has been devoted to the
reconstruction of the photon distribution from realistic detectors
\cite{n,o,p,h1,h2}, and an effective method to reconstruct the
photon distribution starting from on/off photodetection has been
suggested \cite{mogy,mogy2,pcount}, developed \cite{ar}, and
implemented in the last five years
\cite{nos1,nos2,bip_mio1,bip_mio,bip_mio3, moroder}. Convincing
results have been obtained for the reconstruction of the photon
distribution of both single- and bi-partite quantum optical states,
thus showing that an appropriate data processing may turn APD into a
powerful tool for quantum state reconstruction. The technique has
been later extended to the reconstruction of the entire density
matrix, including the off-diagonal elements \cite{full_rho}.
\par
The purpose of this paper is to review the theoretical basis of the on/off method,
and of its experimental implementations, as well as its application to the
reconstruction of the density matrix in different optical regimes.
The paper is structured as follows. In Section \ref{s:basic} we give an overview
on the method focusing on the reconstruction of the photon distribution for single-mode
states. Section \ref{s:bip} describes the extension of the method to the bipartite case
and report experimental implementation and results. In Section \ref{s:rho}
we describe how the method can be extended to achieve the reconstruction of the
whole density matrix and report some recent experimental results about the
reconstruction of the quantum state of light from coherent and pseudo-thermal sources.
Finally, Section \ref{s:out} closes the paper with some concluding remarks and
summarizing future perspectives.
%%%%%%%%%%%%%%%%%%%%%%%%
\section{On/off reconstruction of the photon distribution}
\label{s:basic}
Let us consider a single-mode quantum optical state. All the
accessible information on the state can be obtained using the
Born trace rule on its {\em density matrix} ${\rho}$, which,
in the photon number basis reads as follows
\begin{equation}
\label{dens_mat}
{\rho}=\sum_{n,m=0}^{\infty}\rho_{nm}|n\rangle\langle m|
\end{equation}
In particular, the information about the photon distribution
of the state is given by the diagonal elements $\rho_n\equiv \rho_{nn}$
of the desnsity matrix.
\par
In this Section we are going to show how reconstruction of the
$\rho_n$'s for a general quantum optical state is possible upon exploiting
the set of binary data obtained by a two-level, on/off, detector. We assume
that our state is revealed by a detector like silicon APDs (avalanche
photo-diodes) or a photomultiplier operating in Geiger mode, i.e. a detector
discriminating only between the absence of the presence of the light with
a quantum efficiency $0\leq\eta\leq1$. If we label by $0$ (off) and $1$ (on)
the two possible outcomes, the overall measurement process for this kind
of detector may be described by a two-value \textit{positive
operator-valued measure} (POVM) of the form
\begin{equation}\label{POVM}
    \Pi_{0}(\eta)=\sum_{n=0}^{\infty}(1-\eta)^n|n\rangle\langle n|,
\quad \Pi_1(\eta)=\mbox{I}-\Pi_0(\eta).
\end{equation}
The off probability, $p_0 = {\rm Tr}\left[\rho\,\Pi_{0}(\eta)\right]$,
is thus given by:
\begin{equation}\label{p0p1}
 p_0(\eta)=\sum_{n=0}^{\infty}(1-\eta)^n\rho_n=
    \sum_{n=0}^{\infty}A_n(\eta)\rho_n,
\end{equation}
and the detection probability by $p_1(\eta)=1-p_0(\eta)$.
We assume to have the possibility of varying the quantum efficiency
$\eta$ of our detector and to perform $K$ on/off measurements, each
one with a different $\eta$. The set of experimental data will thus
a sample from the overall distribution
\begin{equation}\label{p0eta}
\mathbf{P_0}\equiv\left\{p_{\mu}:\:\:\:p_{\mu}=
\sum_{n=0}^{\infty}A_{\mu n}\rho_n \:\:\:\:\:\:\:\:\:\:\mu=1,...,K\right\}
\end{equation}
where we have defined $A_{\mu n}\equiv A_n(\eta_{\mu})$.
According to the normalization condition of quantum states we may always
assume $\rho_n\simeq 0$, $\forall n>N$ for some $N$. Eq. (\ref{p0eta}) may be thus
rewritten as a linear system:
\begin{equation}\label{PVrho}
    \mathbf{P_0}=\mathbb{A}\cdot\boldsymbol{\rho}\qquad
\mathbf{\rho}\equiv\left\{\rho_1,\dots,\rho_N\right\}
\end{equation}
where the matrix of coefficients $\mathbb{A}$ is a nonsingular
Vandermonde matrix of order $N+1$, whose coefficients are the geometric
progression $(1-\eta_\mu)^n$. \par
The simplest way to extract the photon distribution from the above
relation is of course via matrix inversion. On the other hand, the form
of the matrix $\mathbb{A}$ is such that numerical inversion with $\mathbf{P_0}$
substituted by the experimental frequencies $\mathbf{F_0}$ is highly inefficient, i.e it
would require a huge number of
experimental runs to avoid large numerical fluctuations. The need of a faster and
statistically more reliable solution thus arises. A closer look to Eq. (\ref{PVrho})
reveals that it represents as a linear positive (LINPOS) statistical model
for the unknowns $\rho_n$ coefficients, which may be effectively solved
by means of a maximum likelihood (ML) approach, i.e. upon finding
the $\rho_n$'s thare are most likely to produce the observed data.
If $n_{\mu}$ is the total number of experimental runs performed with
quantum efficiency $\eta_{\mu}$ and $n_{0\mu}$ the registered number of
off events, then the overall probability, i.e. the likelihood, of the observed
sample is given by ${\cal L}= p_0 (\eta_\mu)^{n_{0\mu}}
(1-p_0 (\eta_\mu))^{n_\mu-n_{0\mu}}$. Upon, using the expectation maximization
(EM) algorithm \cite{EMalg} to maximize the
likelihood functional (actually the log-likelihood $L=\log {\cal L}$),
and imposing the normalization constraint $\sum_n\rho_n=1$, we
arrive at the
iterative formula
\begin{equation}\label{mono_iter}
    \rho_n^{(i+1)}=\rho_n^{(i)}\sum_{\mu=1}^{K}
\left[\frac{A_{\mu n}}{\sum_{\lambda=1}^{K}A_{\lambda n}}\,\frac{f_{\mu}}{p_{\mu}\left[\left\{\rho_n^{(i)}\right\}\right]}\right]
\end{equation}
where $\{\rho_n^{(i)}\}$ is the distribution as reconstructed at the
$i$-th step of the iterative algorithm, $f_{\mu}=n_{0\mu}/n_{\mu}$
is the experimental frequency of the off event with quantum
efficiency $\eta_\mu $, and
$p_{\mu}\left[\left\{\rho_n^{(i)}\right\}\right]$ is the probability
of the off event with quantum efficiency $\eta_\mu$, as evaluated
using the distribution $\{\rho_n^{(i)}\}$ at the $i$-th step
\cite{pcount}.
\par
The formula in Eq. (\ref{mono_iter}) allows one to reconstruct the
photon distribution in terms of experimental off frequencies $f_\mu$
and, of course, of the values of the quantum efficiencies $\eta_\mu$
themselves. Being an iterative algorithm, the need of a measure of
convergence arises. This may be obtained either checking whether it
effectively leads to a maximum for the likelihood functional or,
alternatively, upon introducing an error parameter $\epsilon^{(i)}$,
defined as the distance between the experimental off frequencies and
the corresponding probability reconstructed at the $i$-th step,
\begin{equation}\label{mono_epsilon}
    \epsilon^{(i)}=K^{-1}\sum_{\mu=1}^K\left|f_{\mu}-p_{\mu}\left(\left\{\rho_n^{(i)}\right\}\right)\right|\,.
\end{equation}
In fact, the error $\epsilon^{(i)}$ effectively measures how well
the reconstructed distribution reproduces the observed data. In
turn, we found excellent results upon stopping the iteration number
when the value of $\epsilon^{(i)}$ goes below a certain threshold
quantifying the overall precision of the reconstruction.
\par
In order to assess the accuracy of the method, we consider a
measure of fidelity for the reconstructed photon distribution
in comparison to the theoretical one or expected one $\rho_n^{(th)}$,
as follows
\begin{equation}\label{mono_fidel}
G^{(i)}=\sum_{n=0}^N\sqrt{\rho_n^{(th)}\,\rho_n^{(i)}}\,.
\end{equation}
Several simulated experiments have shown the reliability of the
method, which is also robust against fluctuations in the values of
the quantum efficiencies  $\eta_\mu$. This is a crucial feature for
the experimental implementation of the method, where the values
$\eta_\mu$ are unavoidably determined within some confidence interval,
and may fluctuate during the experimental runs.
\par
\begin{figure}[h!]
\includegraphics[width=0.92\columnwidth]{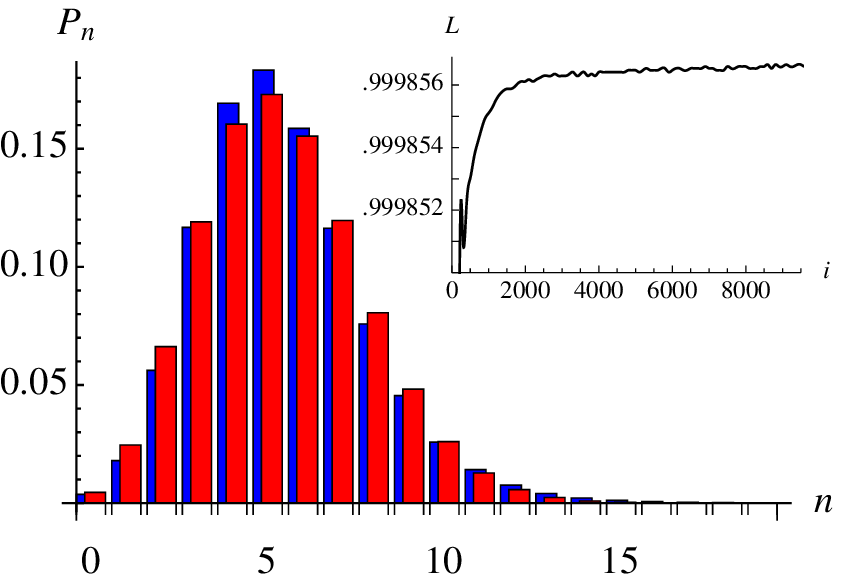}
\includegraphics[width=0.92\columnwidth]{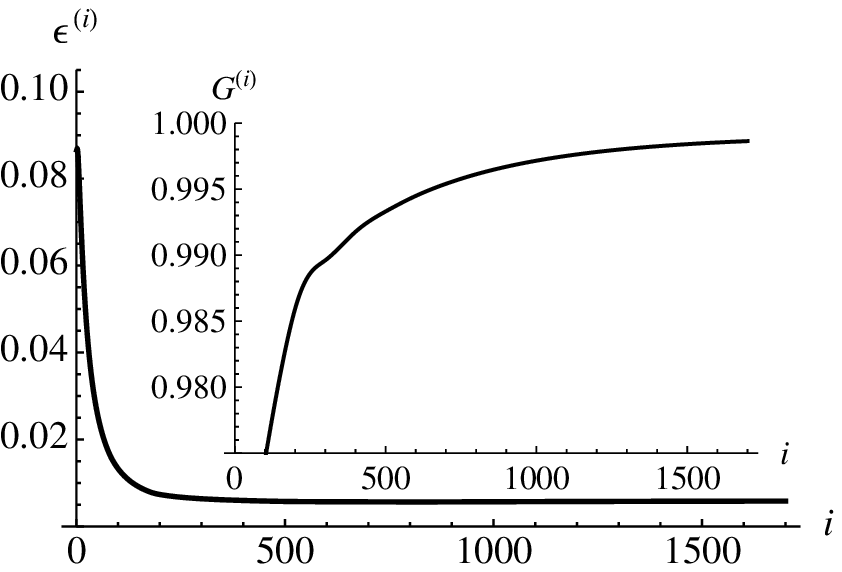}
\caption{On/off reconstruction of single-mode states. The upper panel
shows the reconstruction
of the photon distribution for a weak coherent state (blue) compared to
a Poissonian distribution (red) with the same mean value. The data
corresponds to a coherent state with a mean number of photons
$\left\langle n\right\rangle=5.39$. The inset shows the behaviour of
the -log-likelihood $L$ as a function of the number of steps
of the iterative algorithm. In the lower panel we show the behaviour of the
error parameter $\epsilon^{(i)}$ and of the fidelity $G^{(i)}$ as a function
of the number of steps of the iterative algorithm.} \label{recexample}
\end{figure}

Since the maximum likelihood method provides an asymptotically
unbiased estimator, the confidence interval on the determination
of the element $\rho_n^{(i)}$ can be estimated, for large number
of measurements, in terms of the variance:
\begin{equation}\label{mono_sigma}
    \sigma_n^2=(KF_n)^{-1}
\end{equation}
being $F_n$ the Fisher information:
\begin{equation}\label{mono_fisher}
    F_n=\sum_{\mu=1}^K\frac1{l_\mu}\left(\left.\frac{\partial l_\mu}{\partial\rho_n}\right|_{\rho_n=\rho_n^{(i)}}\right)^2
\end{equation}
where
\begin{equation*}
l_\mu=\frac{\sum_{n=0}^N A_{\mu n}\rho_n}{\sum_{\mu=1}^K\sum_{n=0}^N A_{\mu n}\rho_n}.
\end{equation*}
As a first example of experimental reconstruction, we consider a
weak coherent state generated by a cw He-Ne laser emission and detected
by an APD. The maximum quantum efficiency is given by the nominal
efficiency of the detector, i.e. $\eta=0.66$, whereas the lower values
of $\eta$ needed by the algorithm are obtained by inserting neutral
filters on the optical path of the signal beam. In the upper panel of
Fig.\ref{recexample} we show the reconstructed photon number distribution,
compared to a Poissonian distribution with the same mean value. In turn, a
best fit procedure shows that the reconstructed photon number distribution
is compatible with the one expected for a coherent state with a mean number
of photons equal to $\left\langle n\right\rangle=5.39$. We also show the
convergence of the algorithm: the inset in the upper panel shows
the log-likelihood $L$ as a function of the number of steps of the
iterative formula whereas in the lower panel we report the error parameter
$\epsilon^{(i)}$ and the fidelity $G^{(i)}$ as a function of the number of steps.
At convergence we have fidelity $G\gtrsim 0.995$ between the reconstructed
distribution and the expected Poissonian.
%%%%%%%%
\par
%%%%%%
As a second example, we consider an experiment where single-photon
states are generated by means of a PDC heralding technique
\cite{nos1,nos2}. In this scheme, a pair of correlated photons of
different polarization are produced by pumping a type-II
$\beta$-barium-borate (BBO) crystal with a CW argon ion laser beam
(351 nm) in collinear geometry. After having split the photons of
the pair by means of a polarizing beam splitter, the detection of
one of the two by a silicon avalanche photodiode detector
(SPCM-AQR-15, Perkin Elmer) is used to herald the presence of the
second photon in the other channel, that is a window of 4.9 ns is
opened for detection in the second arm, in correspondence to the detection of
a photon in arm 1. This heralded photon is then measured by another
APD (SPCM-AQR-15, Perkin Elmer) preceded by an iris and an
interference filter (IF) centered at 702 nm (4 nm of FWHM) inserted
with the purpose of reducing the noise due to the stray light. The
overall quantum efficiency of the detection apparatus is $\eta=20\%$
(estimated by Klyshko method \cite{k1pdccs1,k2}), whereas lower
quantum efficiencies are obtained by inserting calibrated neutral
filters (NF) on the optical path of the heralded photon.
\par
The reconstructed photon distribution is excellent agreement with
the expected one: together with a dominant single photon component,
a vacuum, $\rho_0=0.027\pm 0.002$, and double pair $\rho_2=0.019\pm 0.002$
components have been obtained. Those are quantitatively in agreement
with the expectations, due to a small rate of dark counts in the heralding
detector, which trigger the measurement in absence of the heralded photon,
and to the presence of a small multiphoton component in the PDC output.
%%%
Finally, we mention a recent experiments, where the method was applied
to the reconstruction of the photon distribution of one arm of
stimulated PDC, i.e. PDC with a coherent seed in the signal mode
\cite{stimulJMO}. In this case, a generalized version of the reconstruction
algorithm have been developed, which allows us to include constraints on specific
moments, and it has been shown that it provides very good reconstruction in
critical cases.
%%%%%%%%%%%%%%%%%%%%%%%
\section{Extension to the bipartite case}
\label{s:bip}
Multimode states often occur in quantum optical implementations
of quantum information processing, either because of the intrisic
properties of the involved interactions, or for the need of the specific
application. In turn, in this Section, our aim is to present the extension
of our method to the multimode case. In particular, we focus attention to
the bipartite case and present the results of a set of experiments, performed
to reconstruct the joint photon distribution of the two modes exiting a beam splitter
\cite{bip_mio}. In order to generalize the formulas of the previous Section
we assume to deal with a pair of modes that are detected by two on/off
detectors. There are four possible outcomes, which occur with the
following probabilities
\begin{equation}\label{p0011}
    \left\{
    \begin{array}{l}
      p_{00}(\eta) = \sum_{n,k} A_{n}(\eta) A_{k}(\eta) \varrho_{nk}
      \\\\
      p_{01}(\eta) = \sum_{n,k} A_{n}(\eta) [1- A_{k}(\eta)] \varrho_{nk}
      \\\\
      p_{10}(\eta) = \sum_{n,k} [1-A_{n}(\eta)] A_{k}(\eta) \varrho_{nk}
      \\\\
      p_{11}(\eta) = 1- p_{00}(\eta) - p_{10}(\eta) - p_{01}(\eta)
    \end{array}
    \right.
\end{equation}
where:
\begin{equation}\label{varrho}
    \varrho_{nk}=\langle nk | \varrho |nk \rangle
    \qquad |nk\rangle = |n\rangle \otimes |k \rangle
\end{equation}
is the joint photon distribution of the bipartite state, i.e. the diagonal
elements of the four-index two-mode density matrix, and $\eta$ is
the quantum efficiency of the photodetectors, which we assume to be identical.
The above equations provide a relation between the statistics of the
clicks of the two detectors and the actual statistics of photons.
Upon assuming, because of the normalization, tha the elements
$\varrho_{nk}\simeq0$ are negligible beyond some threshold
$\forall\:n,k\geq N$, we may work in a (bipartite) truncated
$(N+1)\times(N+1)$ Hilbert space. Again, if we can
properly change the quantum efficiency of our system in such a way
that $K$ different measurements can be performed (with $K$ different
values $\eta_\mu$, $\mu=1,...,K$, ranging from $\eta_1=\eta_{\min}$
to a maximum value $\eta_{K}=\eta_{\max}$), the whole amount of
on/off detection statistics collected may give enough information to
reconstruct the joint photon distribution of
the bipartite state.
\par
In more details, by re-ordering the diagonal matrix elements according to
the rule
$$\varrho_{nk}\rightarrow q_p \qquad p=1+k+n\,(1+N)$$
we may define the vectors
\begin{align}\label{gq}
\mathbf{g} &=\left(p^{\eta_1}_{00},... ,p^{\eta_K}_{00},
p^{\eta_1}_{01},... ,p^{\eta_K}_{01},p^{\eta_1}_{10},... ,
p^{\eta_K}_{10}\right) \\
\mathbf{q}&=\left(\varrho_{00},\varrho_{01},\varrho_{10},...\right)
\end{align}
and thus summarize the on/off statistics with the compact formula:
$$g_\mu =
\sum_{p=1}^{(N+1)^2} B_{\mu p} q_p \qquad\mu=0,...,3K
$$
that is
\begin{equation}\label{con}
\mathbf{g}=\mathbb{B}\cdot\mathbf{q}
\end{equation}
where we have introduced the matrix $\mathbb{B}$ with entries:
\begin{equation}\label{B_mat}
    [\mathbb{B}]_{\mu p}=
    \left\{
    \begin{array}{l c l}
    A_{\mu n} A_{\mu k} & &\mu=1,..,K \\\\
    A_{\mu n} (1- A_{\mu k}) & & \mu=K+1,..,2K \\\\
    (1 - A_{\mu n}) A_{\mu k} & & \mu=2K+1,..,3K
    \end{array}
    \right.
\end{equation}
where, inverting the transformation rule introduced above, wer have
  $k=(p-1)\mbox{mod}(1+N)$ and $n = (p-1-k)/(1+N)$.
\par
Eq. (\ref{con}) represents a finite statistical
linear model for the positive unknown $q_p$. As for the single-mode case
we may solve the model using a ML approach and, in particular, we may
use EM algorithm to obtain an iterative solution of the maximization
problem
\begin{equation}\label{bip_iter}
    q^{(i+1)}_p =q^{(i)}_p
\sum_{\mu=1}^{3K}\left[\frac{B_{\mu p}}{\sum_\lambda
    B_{\lambda p}} \frac{h_\mu}{g_\mu [\{q^{(i)}_p\}]}\right]\,,
\end{equation}
where $q_p^{(i)}$ denotes the $p$-th element of the reconstructed joint distribution,
$g_{\mu}\left[\left\{q_p^{(i)}\right\}\right]$ are the detection probability reconstructed
at the $i$-the step and $h_{\mu}$ are the experimental frequencies of the
off events, i.e
\begin{align}
    h_{\mu}=\left\{
    \begin{array}{lcl}
    f_{00}=n_{00\mu}/{n_\mu} & & \mu=1,..,K \\
    f_{01}=n_{01\mu}/{n_\mu} & &  \mu=K+1,..,2K \\
    f_{10}=n_{10\mu}/{n_\mu} & & \mu=2K+1,..,3K
    \end{array} \right.
\end{align}
being $n_{01\mu},n_{10\mu},n_{00\mu}$ the number of single and double off
events observed on the whole amount $n_\mu$ of experimental runs performed
with $\eta=\eta_\mu$.
\par
In order to evaluate the confidence interval on the determination of the
element $q_n^{(i)}$ we still use eq. (\ref{mono_sigma}),
but now the Fisher information $F_p$ is
rewritten as
\begin{equation}\label{bip_fisher}
    F_p=\sum_{\mu=1}^{3K}\frac1{d_\mu}\left(\left.
    \frac{\partial d_\mu}{\partial q_p}\right|_{q_p=q_p^{(i)}}\right)^2
\end{equation}
with
\begin{equation*}
    d_\mu=\frac{\sum_{p=1}^{(N+1)^2} B_{\mu p}q_p}{\sum_{\mu=1}^{3K}
\sum_{p=1}^{(N+1)^2} B_{\mu p}q_p}.
\end{equation*}
The analogous of the total error $\epsilon^{(i)}$ of Eq.
(\ref{mono_epsilon}) is given by
$\epsilon^{(i)}=(3K)^{-1}\sum_{\mu=1}^{3K} \left| 
h_\mu-g_\mu\left[\left\{q^{(i)}_p\right\}\right]\right|$,
which measures the distance of the reconstructed off
probabilities from the measured frequencies. Also for the two-mode case,
the algorithm may stopped when $\epsilon^{(i)}$ achieves a
minimum or goes below a certain threshold value.
Finally, the fidelity equation (\ref{mono_fidel}) for the
reconstructed joint photon number distribution may be generalized
$G^{(i)}=\sum_{p=1}^{(N+1)^2}\sqrt{q_p^{(th)}\,q_p^{(i)}}$
giving us the chance to compare the obtained $q_p^{(i)}$ with the
expected ones ($q_p^{(th)}$).
\par
Let us now report the experimental results obtained by applying the
reconstruction method presented above to two different bipartite
states. The first is the state obtained with a single photon state
passing through a beam splitter (BS), whereas the second corresponds
to the splitting of a PDC single branch. In both cases, which corresponds
to very different optical regimes, the algorithm provides good reconstruction
of the joint photon distribution \cite{bip_mio,bip_mio3}.
\par
\begin{figure}[h!]
\includegraphics[width=0.92\columnwidth]{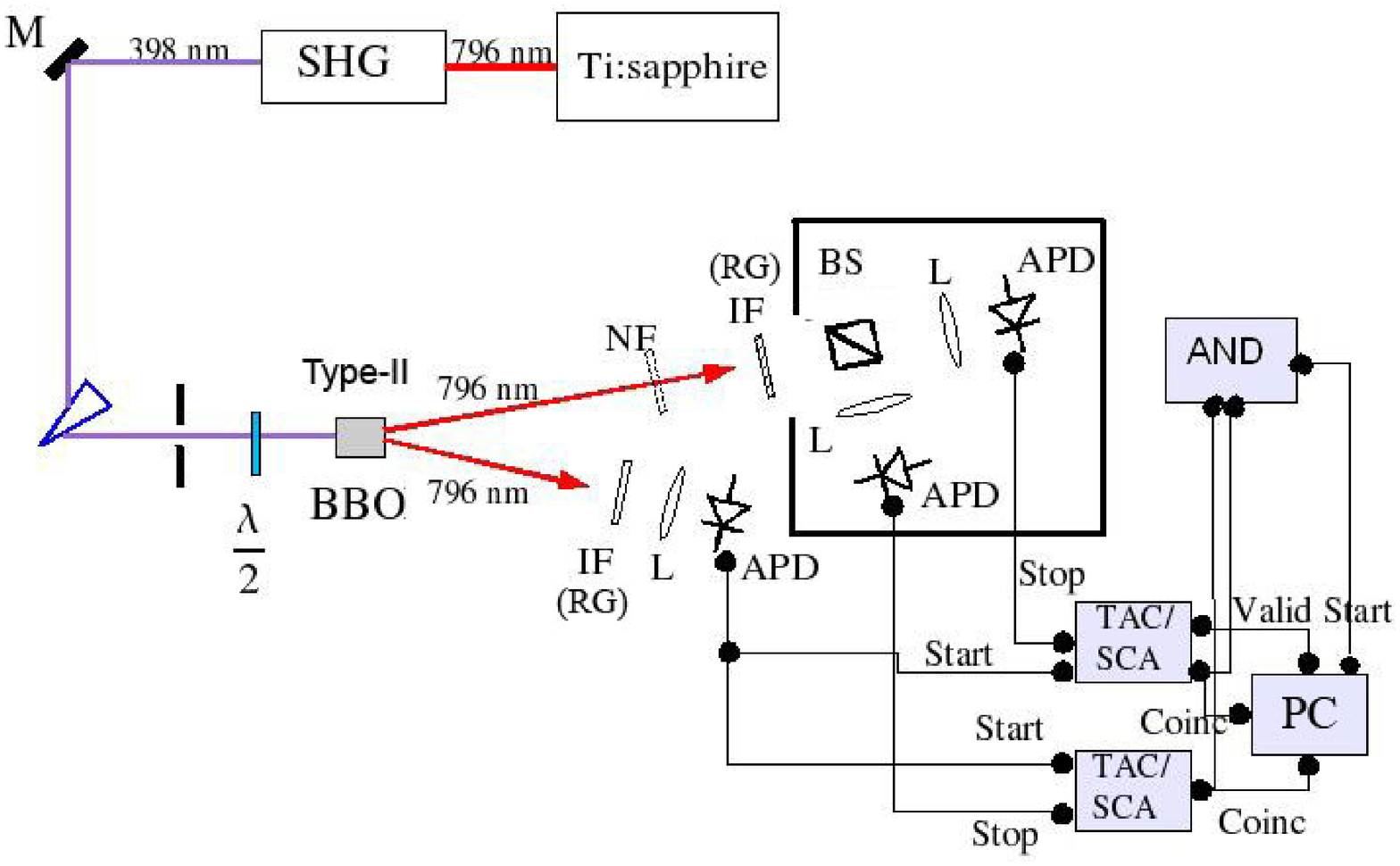}
\caption{On/off reconstruction of two-mode states. Schematic diagram of
the experimental setup realizing the type-II PDC heralded photon source
used to generate a two-mode superposition of a single-photon state
with the vacuum. The idler photon is addressed to an IF (RG)
filter, collected and sent to APD1, opening a coincidence window in the
TAC modules; the signal goes through the NF and the IF (RG) filters, and
then is split by the BS, whose outputs are collected and sent to  APD2
and APD3 to close the coincidence windows. The output of the two
TACs is also sent to an AND logical gate whose outputs gives the number
of double coincidences.} \label{bip_setup1.eps}
\end{figure}

In our first experimental setup, see Fig. \ref{bip_setup1.eps},
\cite{bip_mio}, a 0.2 W, 398 nm pulsed (with 200 fs pulses and 70
MHz repetition rate) laser pump have been generated by second harmonic of
a Ti:Sapphire laser at 796 nm and then injected into a
$5\times5\times1$ mm type-II BBO crystal, this leading to the
generation of entangled photon pairs by parametric downconversion.
The detection of a photon on one of two correlated branches of degenerate
PDC emission is used as trigger to herald the presence of the correlated
photon in the other direction.
%%%
%%%
\par
\begin{figure}[h!]
\includegraphics[width=0.92\columnwidth]{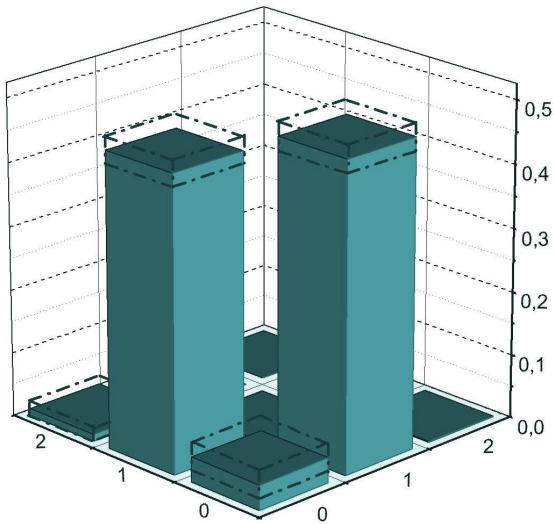}
\caption{On/off reconstruction of two-mode states. The plot shows the
reconstructed $\varrho_{nk}$ entries of the joint photon
distribution of our $\left|\psi_{BS}\right\rangle$ for the setup
with 50\%-50\% BS and 10 nm FWHM IF.}
\label{IF10(5050)bis1.eps}
\end{figure}

The idler photon is addressed to an optical filter (a narrow
band interference filter or a red glass, as will be specified
later), collected by a lens and finally sent to a silicon APD (APD1).
The corresponding signal is properly filtered (with the
same filter as the idler photon) and then sent into a beam
splitter (BS), separating in two the optical path of the photon and
thus generating a bipartite state, that is the nonlocal superposition
of a single photon state and the vacuum in the two arms
\begin{equation}\label{BS_psi}
    \left|\psi_{BS}\right\rangle=\sqrt\tau|0\rangle
|1\rangle+\sqrt{1-\tau}|1\rangle |0\rangle
\end{equation}
where $\tau$ is the BS transmittance. The BS is followed, on both
output arms, by a collection/detection apparatus, denoted by APD2
and APD3, of the same type as APD1 (all the detectors were Perkin
Elmer SPCM-AQR-15 silicon APDs). The proper set of quantum
efficiencies are obtained by inserting, before the BS, several
neutral filters (NF) of different transmittance, calibrated by
measuring the ratio between the counting rates on D2 and D3 with the
filter inserted and without it. In correspondence of the detection
of a photon in arm 1, a coincidence window is opened on both
detectors on arm 2: this may be obtained by sending the output of D1
as Start to two Time-to-Amplitude Converters (TAC) that received the
detector signal of D2 and D3 as stop. The 20 ns coincidence window
has been set such to avoid spurious coincidences with PDC photons
belonging to the following pulse (we remind that the repetition rate
of the laser is 70 MHz). The TAC outputs are then addressed to the
computer and to an AND logical gate in order to reveal coincidences
between them; its output is also collected via computer, together
with one TAC's valid start, giving us the total number of opened
coincidence windows. These four data sets allow evaluating the
frequencies $f_{00}$, $f_{01}$, $f_{10}$, $f_{11}$, needed for the
reconstruction of the photon statistics. The background is estimated
and subtracted by measuring the TACs and AND outputs out of the
window triggered by APD1 detection. The maximum quantum efficiency
is evaluated as the ratio between the sum of coincidences in APD2
and APD3 and the counting rate on APD1 without the insertion of any
NF \cite{pdccs,pdccs1}. In order to verify the method in different
conditions we have also considered four alternatives given by the
combination of a balanced (50\% - 50\%) or unbalanced (40\% - 60\%)
BSs with two optical filters sets, either large band red glass
filters (RG) with cut-off wavelength at 750 nm, or interference
filters (IF) with peak wavelength at 796 nm and a 10 nm FWHM. The
first test has been performed with the 50\%-50\% BS and the
interference filters, for which we have collected data for $K=33$
different quantum efficiencies. Elaborating these data with our
reconstruction algorithm within a $3\times3$ Hilbert space choice
($N=2$) lead to the reconstructed joint photon distribution shown in
Fig. \ref{IF10(5050)bis1.eps}: here we can appreciate how the only
relevant entries are $\varrho_{01}$ and $\varrho_{10}$ (single
photon transmitted or reflected by the beam splitter), in good
agreement with the inserted BS ratio.
%%%
%%%
\par
There is also a small nonzero vacuum component $\varrho_{00}$ which is due to a
non perfect background evaluation and subtraction in the experimental data and perhaps
to some light absorption in the system caused by the BS cube.
An additional test has been performed with the same BS but with different
optical filters (red glass (RG) filters with
$\lambda_{cut-off}=750$ nm) in order to explore a different range
of frequencies.
The overall quantum efficiency in this experiment is $\eta=0.088$, and the
acquisition has been repeated with $K=33$ decreasing values. Effective
reconstruction of the photon distribution has been obtained also in this case.
Then we replaced the 50\%-50\% BS cube with an unbalanced 60\%-40\%
BS plate, maintaining the RG large band filters: with this setup we
have an overall quantum efficiency $\eta=0.0123$, and have performed
$K=41$ data collection. The difference between the transmitted and reflected
branch of the BS was evident and in agreement with the known beam
splitter ratio, with the only nonzero $\varrho_{nk}$ being $\varrho_{10}$ and
$\varrho_{01}$. Finally, for the last test we have replaced the RG filters
with the previous IF ones and obtain a similar results, i.e. excellent reconstruction
of the joint photon distribution.
\par
As a second example we consider the two-mode state obtained by inserting
a single branch of type-I PDC emission without triggering, which corresponds
to a multi-thermal multi-photon state \cite{Mandel}, into a balanced beam
splitter. In a setup similar to the previous one, \cite{bip_mio}, we have
generated PDC light by means of a $5\times5\times5$ mm Type-I BBO crystal
pumped by a Q-switched (triplicated to 355 nm) Nd:Yag laser with 5 ns pulses,
power up to 200 mJ per pulse and 10 Hz repetition rate. Because of the very
high power of the pump beam, a state with a large number of photon is generated
for each pulse. We have therefore attenuated (by using 1 nm FWHM IF and neutral
filters) the multithermal state before sending it to the BS and then detecting
both the outgoing beams.
Again, different quantum efficiencies are obtained by inserting (before the BS)
Schott neutral filters, whose calibration is obtained by measuring the power of
a diode laser before and after them, with the calibration laser injected in the
same point as the PDC in order to minimize the effect due to eventual non
homogeneous NF filters. The coincidence scheme is realized sending two Q-switch
triggered pulses to two TAC modules as start inputs, and the APDs outputs as stop.
Then, having set properly the 20 ns coincidence window, we address the two TAC
outputs to the AND logic port, and the valid stops to the counting modules
(together with one TAC's valid start and the AND output). The background is
estimated and subtracted in the same way as in the previous experiment.
The maximum quantum efficiency in this experiment has been
evaluated by multiplying the APDs nominal quantum efficiencies, the
IF peak transmittance and the fiber couplers effective efficiency,
measured with the diode laser leading to $\eta=0.25$.
%%%
The expected on/off joint statistics for this optical state is:
\begin{equation}\label{multi-stat}
    \left\{
        \begin{array}{l}
         p^\eta_{00}= [M(M+\eta N_{ave})^{-1}]^{M} \\\\
         p^\eta_{01}= [M(M+\eta \tau N_{ave})^{-1}]^M-p^\eta_{00}\\\\
         p^\eta_{10}= \{M[M+\eta (1-\tau)N_{ave}]^{-1}\}^M-p^\eta_{00}
        \end{array}
    \right.
\end{equation}
where $N_{ave}$ is the average number of photons, $M$ the number of
propagation modes and $\tau$ the BS transmittance.
The reconstructed joint photon statistics (now upon a $17\times17$
truncated Hilbert space) have been compared with the expected two-mode
multithermal distribution
\begin{equation}\label{multith_prob}
    \varrho_{nm}=\frac{(n+m+M-1)!}{n!m!(M-1)!}\cdot
    \frac{\left(1+\frac{N_{ave}}M\right)^{-M}}{\left(1+\frac M{N_{ave}}\right)^{m+n}}\,
\end{equation}
and the calculated  fidelity is larger than $G\simeq 0.99$ for
$i\geq2000$. Overall, this experiment confirms of the reliability
of our method, even when it is dealing with a larger Hilbert space and
more intense beams, corresponding to quantum states with a larger number
of photons.
%%%%%%%%%%%%%%%
\section{Full state reconstruction by on/off photodetection}
\label{s:rho} The method described in the previous Sections may
provide the complete information on a quantum optical state when the
density matrix of such a state is diagonal, i.e. all the
off-diagonal elements are equal to zero. Of course, in real systems,
it often occurs that also off-diagonal elements are relevant and a
question arises on whether the on/off method may be generalized to
obtain a more general procedure providing the reconstruction of the
whole density matrix. In turn, the answer is positive, and basically
require to supplement the on/off method with some additional phase
information. Before entering into the details of our implementation
let us consider a generic single-mode state, mixed with a strong
coherent state, from now on the local oscillator (LO), in an
unbalanced beam-splitter, i.e. high transmittance and low
reflectance. In this case then the transmitted mode is {\em
displaced} \cite{bs96}, i.e. it is equivalent to the signal mode
shifted by a displacement operator $D(\alpha)=\exp(\alpha
a^\dagger-\alpha^*a)$, where $a$ ($a^\dagger$) is the photon
destruction (creation) operator associated to the signal mode and
$\alpha=\left|\alpha\right|e^{i\varphi}$ is the local oscillator
field amplitude rescaled by the BS reflectance. If one measures the
photon distribution of the transmitted beam, this corresponds to
measuring the displaced Fock-state probability distribution of the
original signal, i.e.
\begin{equation}
\label{eq:statdisplacement}
p_n(\alpha)=\trabra{n,\alpha}{\rho}\traket{n,\alpha}
\end{equation}
where $\traket{n,\alpha}\equiv D(\alpha)\traket{n}$
are the displaced Fock states and
${\rho}$ is the signal mode density operator. Note that
$p_n$ is now a function of the displacement $\alpha$ also.
\par
The on/off state reconstruction method is, in turn, based on the
above equation \cite{opa:PRA:97}. In fact, upon truncating the Hilbert
space at dimension $n_0$ we can expand eq. \ref{eq:statdisplacement}
as follows
\begin{equation}
\label{eq:statdisplacement2}
p_n(\alpha)=\sum_{k,m=0}^{n_0}\left\langle n,\alpha
\traket{k}\trabra{k}{\rho}\traket{m}\trabra{m}
n,\alpha\right\rangle
\end{equation}
Expressing the displaced Fock states $\traket{n,\alpha}$ in the
ordinary Fock basis, one obtains\cite{opa:PRA:97}:
\begin{align}
\label{eq:expandedpn}
p_n(\alpha)&= e^{-\left|\alpha\right|^2}n! \sum_{k,m=0}^{n_0}\sqrt{k!m!}
\trabra{k}{\rho}\traket{m}\times\nonumber\\
&\times\sum_{j=0}^{\bar j}\sum_{l=0}^{\bar l}\frac{(-1)^{j+l}
\left|\alpha\right|^{m+k+2(n-j-l)}e^{i(m-k)\varphi}}
{j!(n-j)!(k-j)!l!(n-l)!(m-l)!}
\end{align}
where $\bar j \equiv min\{n,k\}$ and $\bar l \equiv min\{n,m\}$.
Now, we assume $|\alpha |$ fixed, so that for any value of
$\left|\alpha\right|$, $p_n(\alpha)$ can be
regarded as a function of $\varphi$ and expanded in a Fourier
series, the general component being:
\begin{align}
\label{eq:fourier}
p_n^{(s)}(\left|\alpha\right|)&=
\frac{1}{2\pi}\int_0^{2\pi}{p_n(\alpha)e^{is\varphi}d\varphi}\nonumber\\
&=\sum_{m=0}^{n_0-s}G_{n,m}^{(s)}(\left|\alpha\right|)\trabra{m+s}{\rho}\traket{m}
\end{align}
with:
\begin{align}
\label{eq:g}
G_{n,m}^{(s)}&(\left|\alpha\right|)=e^{-\left|\alpha\right|^2}n!
\sqrt{m!(m+s)!}\times\nonumber\\
&\times\sum_{j=0}^{\tilde j}\sum_{l=0}^{\bar l}\frac{(-1)^{j+l}
\left|\alpha\right|^{2(m+n-j-l)+s}}{j!(n-j)!(m+s-j)!l!(n-l)!(m-l)!}
\end{align}
where $\tilde j \equiv min\{n,m+s\}$.
We notice that $p_n^{(s)}(\left|\alpha\right|)$ is related to the
density matrix element whose row and column indices differ by $s$.
If the photon number distribution $p_n(\alpha)$ is measured for
$n=0,1,\dots,N$, with $N\geq n_0$, then Eq. (\ref{eq:fourier}) represents,
for each fixed value of $s$, a system of $(N+1)$ linear equations
connecting the $(N+1)$ measured quantities $p_n^{(s)}$ to
$(n_0+1-s)$ unknown density matrix elements.
This system is clearly overdetermined, so it can be inverted using
the least squares method in order to obtain the density matrix
elements from the measured probabilities.
The reconstructed off-diagonal density matrix elements can thus be
obtained as:
\begin{equation}
\label{eq:offd}
\trabra{m+s}{\rho}_{rec}\traket{m}=\sum_{n=0}^NF_{n,m}^{(s)}
(\left|\alpha\right|)p_n^{(s)}(\left|\alpha\right|)
\end{equation}
where
\begin{equation}
F_{n,m}^{(s)}(\left|\alpha\right|)=\{[G_{n,m}^{(s)}
(\left|\alpha\right|)]^TG_{n,m}^{(s)}(\left|\alpha\right|)\}^{-1}
[G_{n,m}^{(s)}(\left|\alpha\right|)]^T
\end{equation}
is the generalized Moore-Penrose inverse of $G$.
The $F$ matrix satisfies the condition:
\begin{equation}
\sum_{n=0}^N F_{m',n}^{(s)}(\left|\alpha\right|)
G_{n,m}^{(s)}(\left|\alpha\right|)=\delta_{m,m'}
\end{equation}
for $m,m'=0,1,\dots,n_0-s$, so that from the exact probabilities the
correct density matrix elements are found
(${\rho}_{rec}\equiv{\rho}$). Furthermore, the least squares
method ensures that the $^{*}p_n^ {(s)}$ calculated from
${\rho}_{rec}$ according to eq. (\ref{eq:fourier}) best fits the
measured quantities such that $
\sum_{m=0}^N(^{*}p_n^{(s)}-p_n^{(s)})^2$
is minimized.
In conclusion, upon combining Eqs. (\ref{eq:offd}) and (\ref{eq:fourier}), we
find out the formula for the direct sampling of the density matrix
from the measured photon number distribution of the displaced state\cite{opa:PRA:97}:
\begin{equation}
\label{eq:directsampling}
\trabra{m+s}{\rho}_{rec}\traket{m}=\frac{1}{2\pi}\sum_{n=0}^N
\int{F_{n,m}^{(s)}(\left|\alpha\right|)e^{is\varphi}p_n(\alpha)d\varphi}
\end{equation}
It is worth mentioning that this method, given that the
$p_n(\alpha)$ are known,  requires only the value of $\varphi$ to be
varied.
\par
Let us now extend our method to the case of imperfect photodetection
with nonunit quantum efficiency $\eta$. The measured photodistribution
$P_k(\alpha)$ is related to the actual photon number distribution $p_n(\alpha)$
by a Bernoullian convolution
\begin{equation}
P_k(\alpha)= \sum_{n=0}^\infty M_{k,n}(\eta)p_n(\alpha)
\end{equation}
where $M_{k,n}(\eta)$ is:
\begin{equation}
M_{k,n}(\eta)= \left\{
    \begin{array}{ll}
    \binom{n}{k}\eta^k(1-\eta)^{n-k} &\quad k\leq n \\\\
    0 &\quad k>n
    \end{array}
    \right.\,.
\end{equation}
The analogue of Eq.(\ref{eq:fourier}) for
the measured quantities $P_n^{(s)}(\left|\alpha\right|)$
is given by
\begin{equation}
\label{eq:fourier2}
P_n^{(s)}(\left|\alpha\right|)=\sum_{m=0}^{n_0-s}G_{n,m}^{(s)}
(\left|\alpha\right|,\eta)\trabra{m+s}{\rho}\traket{m},
\end{equation}
where the new matrices
\begin{equation}
G_{n,m}^{(s)}(\left|\alpha\right|,\eta)=\sum_{k=0}^\infty
M_{n,k}(\eta) G_{k,m}^{(s)}(\left|\alpha\right|)\,,
\end{equation}
are obtained from the $G_{k,m}^{(s)}(\left|\alpha\right|)$,
as defined in Eq. (\ref{eq:g}), and can be inverted in the same way
described above to obtain some matrices $F_{m,n}^{(s)}(\left|\alpha\right|,\eta)$
to be used, as in Eq. (\ref{eq:directsampling}), to reconstruct the whole density
matrix.
\par
The key element to realize the above method and, in turn, to achieve the
reconstruction of the off-diagonal elements of the density matrix, is an
interferometric setup where the signal mode and the local oscillator are
mixed in an unbalanced beam splitter. In practice, a source beam in sent
to a Mach-Zehnder interferometer in which the part reflected by the
first beam splitter is taken as the signal, while the transmitted
portion is regarded as the local oscillator. The two modes are then
mixed by the second beam splitter, which is effectively used to perform
a sort of unbalanced homodyning. Upon changing the length of the optical
path in the one arm of the inteferometer, one may tune the relative
phase between the signal mode and the local oscillator.  In this way,
one may measure the distributions $p_n(\alpha )$ at fixed $|\alpha |$
and for different phases. Besides, upon performing some action on the
signal mode, one may also apply the reconstruction method to different
input states.  In the following ,we report the results for the
reconstruction of a coherent state and for a thermal state obtained by
inserting a rotating glass plate in the path of the signal beam.
\par
\begin{figure}[h!]
\includegraphics[width=0.92\columnwidth]{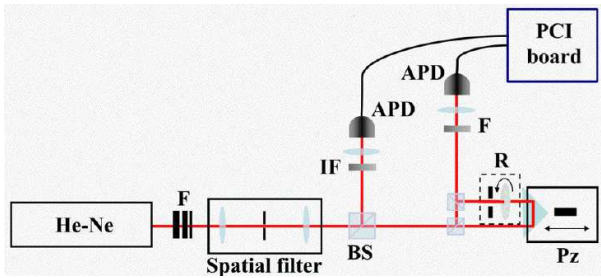}
\caption{{Setup for the reconstruction of the density matrix for a
coherent and a pseudo-thermal state. The emission of a He-Ne laser
($\lambda=632.8$ nm) is lowered to single photon regime by neutral
filters. A spatial filter realized by two converging lenses and a $100$
$\mu$m diameter-wide pinhole purifies the shape of the signal and allows
to select a single spatial mode.  A beam-splitter reflects part of the
beam to a control detector used to monitor the laser amplitude
fluctuations, while the remaining part is sent to the interferometer.
The phase between the "short" and "long" paths  in the interferometer
can be changed by driving the position of the reflecting prism by means
of a PI piezo-movement system. A set of variable neutral filters allows
to collect photons for different values of the quantum efficiency.  The
element in the dotted box is a rotating glass plate which is inserted in
the setup only for the generation of the pseudo-thermal state.   The
detectors used are Perkin-Elmer Single Photon Avalanche
Photodiode(SPCM-AQR) gated by a 20 ns wide time window with (repetition
rate $=$ $200$ kHz). A single run consists of 5 repetitions of 4 seconds
acquisitions and events are recorded by a NI-6602 PCI counting module. }
\label{figg:setup}}
\end{figure}

In our set-up \cite{full_rho}, see Fig.\ref{figg:setup},the
output of a He-Ne laser ($\lambda=632.8$ nm) is lowered to the single
photon regime by neutral filters. The spatial profile of the signal is
the purified from non-Gaussian components by a spatial filter realized
by two converging lenses and a $100$ $\mu$m diameter wide pinhole. An iris
just after the pinhole ensures the selection of a single Gaussian spatial
mode. The laser cavity is also preserved by back-reflections, which may
cause instability, by means of an optical isolator consisting in a Faraday
rotator between two polarizers. The second polarizer (say, B) angle is shifted
by $45^\circ$ with respect to the first one (A), so that the light
transmitted by the latter, whose polarization is rotated by $45^\circ$ by
the Faraday rotator, is all transmitted also by the second. Since the polarization
rotation by Faraday effect is in the same direction regardless of the laser
propagation direction, any back-reflected light passing through B, would
suffer another $45^\circ$ rotation before reaching a polarizer and would
thus be stopped, being orthogonal to the polarizer angle.
%%%%%%%%%%%%%%%%
%%%%%%%%%%%%%%%%%%%%%
\par
After a beam-splitter, part of the beam is addressed to a control
detector in order to monitor the laser amplitude fluctuations, while the
remaining part is sent to the interferometer, its main structure
consisting in a single invar block custom designed and developed at
INRIM. A  PI piezo-movement system allows to change the phase between
the two paths by driving the position of the reflecting prism with
nanometric resolution and high stability. For each position of the
prism, the off events are collected for different sets of neutral
filters, and thus, for different quantum efficiencies. The detector,  a
Perkin-Elmer Single Photon Avalanche Photodiode (SPCM-AQR), is gated by
a 20 ns wide time window with a repetition rate of $200$ kHz. In order
to obtain a reasonable statistics, a single run consists of 5
repetitions of 4 second acquisitions. Events are recorded by a NI-6602
PCI counting module.  In this case, since all the attenuations in front
of the detectors can be included in the generation of the state, the
overall maximum quantum efficiency is assumed to be $0.66$, as the
nominal efficiency declared by the manufacturer data-sheet of the
photodetectors. As said before, in another set of measurements, a
rotating glass plate was inserted in the path of the signal in order to
reconstruct a pseudo-thermal state.  
%%%%%%%%%%
\begin{figure}[h!]
\includegraphics[width=0.92\columnwidth]{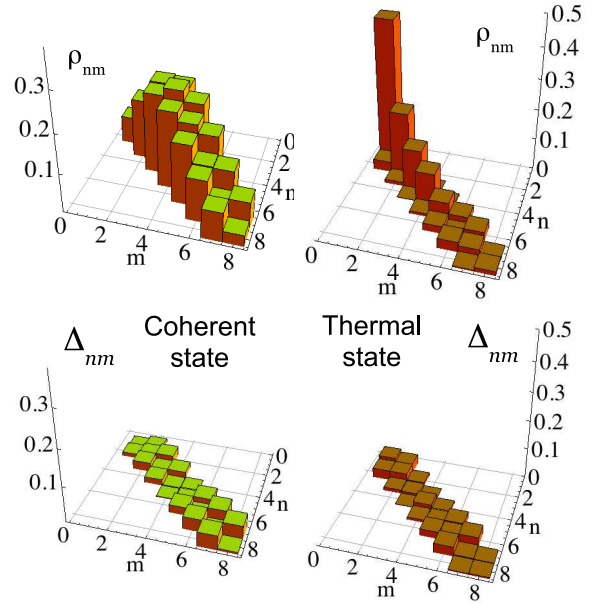}
\caption{State reconstruction by phase-modulation and
on/off measurements. In the upper plots
we report the reconstructed density matrix in the Fock
representation (diagonal and subdiagonal elements) for the signal
excited in a coherent state with 
real amplitude $z\simeq1.8$ (left) and a thermal state with average number 
of photons equal to $n_{th}\simeq1.4$ (right). 
In the lower plots we show the absolute difference $\Delta_{nm}=|\varrho^{exp}_{nm}
-\varrho^{th}_{nm}|$ between reconstructed and theoretical values
of the density matrix elements for the signal excited in a 
coherent (left) or a thermal (right) state.\label{f:err}}
\end{figure}
%%%%%%%%%%
\par
\begin{figure}[h!]
\includegraphics[width=0.92\columnwidth]{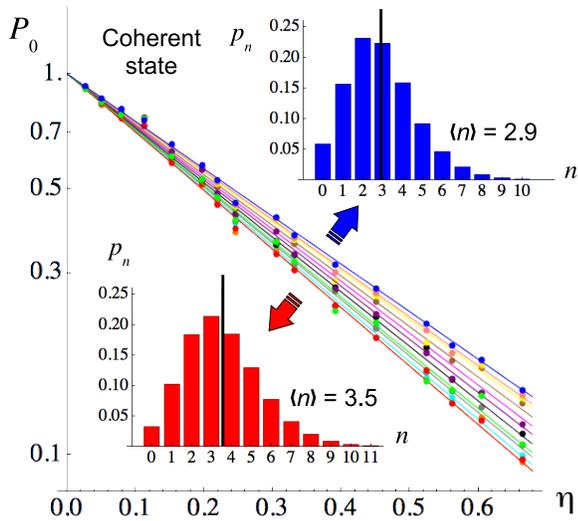}
\caption{State reconstruction by phase-modulation and
on/off measurements. In the main plot we report the off
frequencies as a function of the quantum efficiency as obtained when the
signal under investigation is a coherent state and for different
phase-shifts. The two insets show the reconstructed photon distributions
for the two phase-modulated versions of the signal corresponding to
maximum and minimum visibility at the output of the Mach-Zehnder
interferometer.  The vertical black bars denote the mean value of the
photon number for the two distributions, $\langle a^\dag a\rangle = 3.5$  and
$\langle a^\dag a\rangle = 2.9$.\label{figg:preresults2}}
\end{figure}

In Fig. \ref{f:err} we report the reconstructed density matrix in the 
Fock representation (diagonal and subdiagonal) for a coherent state with 
real amplitude $z\simeq1.8$ and a thermal state with average number of photons 
equal to $n_{th}\simeq1.4$. As it is apparent from the plots, the off-diagonal 
elements are correctly reproduced in both cases despite the limited visibility.  
Here the raw data are frequencies of the off event (see Fig. ref{figg:preresults2}) 
as a function of the detector efficiency, taken at different phase 
modulations $\phi$, whereas the intermediate step corresponds to the 
reconstruction of the photon distribution for the phase-modulated signals. 
In Fig. \ref{figg:preresults2} we
report the frequencies of the off events and the reconstructed photon
distributions at the minimum and maximum of the interference fringes.
In our experiments
we used $N_\phi=12$ and $|\alpha|^2=0.01$ for the coherent state and
$|\alpha|^2=1.77$ for the thermal state.  The use of a larger $N_\phi$
would allow the reliable reconstruction of far off-diagonal elements,
which has not been not possible with the present configuration.  
Work along these lines is in progress and results will be reported
elsewhere.
%%%%%%%%%%
%%%%%%%%%%
\par
The evaluation of uncertainties on the reconstructed states involves the
contributions of experimental fluctuations of on/off frequencies as well
as the statistical fluctuations connected with photon-number
reconstruction.  It has been argued \cite{KER97,KER99} that fluctuations
involved in the reconstruction of the photon distribution may generally
result in substantial limitations in the information obtainable on the
quantum state, e.g. in the case of multipeaked distributions \cite{zz}.
For our purposes this implies that neither large displacement amplitudes
may be employed, nor states with large field and/or energy may be
reliably reconstructed, although the mean values of the fields measured
here are definitely non-negligible.  On the other hand, for the relevant
regime of weak field or low energy, observables characterizing the
quantum state can be safely evaluated. In our experiments, the absolute
errors $\Delta_{nm}=|\varrho^{exp}_{nm}-\varrho^{th}_{nm}|$ on the
reconstruction of the density matrix in the Fock basis are reported in
Fig. \ref{f:err}. Notice also that any uncertainty in the nominal
efficiency of the involved photodetectors does not substantially affect
the reconstruction in view of the robustness of the method to this kind
of fluctuations \cite{pcount}.
%%%%%%%%%%%%%%%%%
\section{Conclusions and outlooks}
\label{s:out}
In this paper we gave a panorama of the status of the art about the
characterization of optical states by means of on/off photodetection.
We have briefly reviewed the theoretical basis of the on/off method, and
have reported the main recent experimental results for the
reconstruction of the diagonal elements of single- and two-mode states,
as well as for the reconstruction of the full density matrix of
single-mode states. The on/off reconstruction method has been now tested
in several optical regimes and it proved both effective and
statistically reliable. This prompts to further applications of the
scheme as a tool for the characterization at quantum level. In fact, we
are currently going to apply the on/off method as an advanced
characterization of detectors, i.e. for the reconstruction of their
probability operator-valued mesaures (POVMs).
%%%%%%%%%%%%%%%%%%%
\section*{Acknowledgements}
This work has been supported by EU project QuCandela, by Compagnia di
San Paolo, by the MIUR projects PRIN 2007FYETBY and PRIN2005023443, by
Regione Piemonte (E14) and by the CNR-CNISM agreement. MGAP thanks
Zdenek Hradil, Maria Bondani, and Simone Cialdi for useful 
discussions.
%%%%%%%%%%%%%%%%%%%%%%%%%%%%%%%%%%%%%%%%%%%%%%%

\end{document}